# Neutron background measurements at China Jinping underground laboratory with a Bonner Multi-sphere Spectrometer

Qingdong[a,b] Hu, Hao Ma[a,b], Zhi Zeng[a,b,*], Jianping Cheng[a,b], Yunhua Chen[c], Shenming He[c], Junli Li[a,b], Manbin Shen[c], Shiyong Wu[c], Qian Yue[a,b], Jianfeng Yue[c], Hui Zhang[a,b]

[a] Department of Engineering Physics, Tsinghua University, Beijing 100084, China
[b] Key Laboratory of Particle & Radiation Imaging (Tsinghua University), Ministry of Education, China
[c] Yalong River Hydropower Development Company, Chengdu 610051, China

**Abstract**

The neutron background spectrum from thermal neutron to 20 MeV fast neutron was measured at the first experimental hall of China Jinping underground laboratory with a Bonner multi-sphere spectrometer. The measurement system was validated by a $^{252}$Cf source and inconformity was corrected. Due to micro charge discharge, the dataset was screened and background from the steel of the detectors was estimated by MC simulation. Based on genetic algorithm we obtained the energy distribution of the neutron and the total flux of neutron was $(2.69\pm1.02) \times 10^{-5}$ cm$^{-2}$s$^{-1}$.

## 1. Introduction

China JinPing underground Laboratory (CJPL) is located in the center of a 17.9 km long traffic tunnel through Jinping Mountain in Sichuan Province, Southwest China. With an overburden of 2400 m rocks, it is the deepest underground laboratory in the world and an ideal location for rare event physical experiments, e.g. dark matter search. The first experimental hall of CJPL, called CJPL-1 was built in summer 2010[1], in which two dark matter experiments, i.e. CDEX[2] and PandaX[3] are being operated. Precise knowledge of the background induced by cosmic rays, gammas and neutrons originating from surroundings is important for these experiments to estimate their sensitivity.

---

[*] Corresponding author at: Department of Engineering Physics, Tsinghua University, Beijing 100084, China.
E-mail address: zengzhi@tsinghua.edu.cn (Zhi Zeng).

Neutrons are major background for dark matter experiments, especially the similar recoil signals induced by neutrons and dark matter. In the underground laboratories, neutrons are from muon induced secondary particles, spontaneous fission and (alpha, n) reactions. While the overburden of rock at CJPL is deep, the cosmic ray muon flux in the CJPL is measured to be[4] $(2.0±0.4) \times 10^{-10}/(cm^2 \cdot s)$ and the muon induced neutrons is $8.37 \times 10^{-11}$ $cm^{-2}s^{-1}$ by Monte Carlo simulation[5] To detect neutron spectrum in the underground laboratory, liquid scintillator is widely applied to obtain fast neutron such as Modane[6] and Boulby[7] underground laboratories. While proportional counters is often used to obtain thermal neutron spectrum such as in the Gran Sasso[8]. The flux of thermal neutron at CJPL-1 was detected by a $^3$He proportional tube and a $^4$He proportional tube as the detector [9]. But the research about neutron spectrum from thermal neutrons to fast neutrons is few. And the spectra is important to predict background and design appropriate shielding. Bonner multi-sphere spectrometer is widely used [10] to determine neutron spectrum because of broad energy and isotropic response. The neutron spectrum on the ground has been measured by a Bonner multi-sphere spectrometer [11] and the spectrometer is applied to obtain the neutron spectrum in the CJPL-1.

## 2. Method and experiment

### 2.1 Instrument and setup

The Bonner multi-sphere spectrometer in this study is composed of eight helium-3 spherical proportional counters and seven different size moderators made in polyethylene. Unfolding software of the spectrometer is coded based on genetic algorithm [12]. The response functions of our spectrometer can be found in this paper [12].

Table 1 The Bonner sphere spectrometer system used for this measurement

| Name of | Radius of | Name of counter |
| --- | --- | --- |

| proportional counter | moderator | with moderator |
|---|---|---|
| D-0 | 0 mm | DR-0 |
| D-1 | 50 mm | DR-50 |
| D-2 | 60 mm | DR-60 |
| D-3 | 70 mm | DR-70 |
| D-4 | 80 mm | DR-80 |
| D-5 | 100 mm | DR-100 |
| D-6 | 120 mm | DR-120 |
| D-7 | 150 mm | DR-150 |

The Bonner sphere spectrometer mentioned above is installed at CJPL-1 and the spectrometer is operated about one year. The following figure shows the measuring environment. An initial dataset is acquired and will be analyzed in the following part.

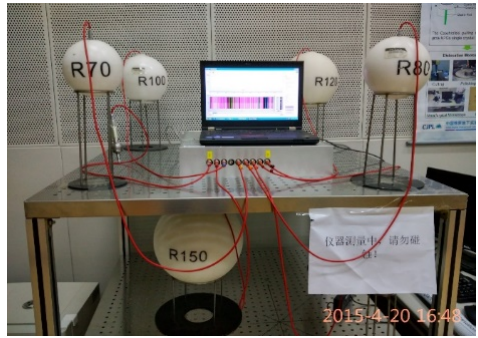

Figure 1 Experimental setup at CJPL-1

## 2.2 Validation of spectrometer

A $^{252}$Cf neutron source was used to validate the reliability of the spectrometer. The $^{252}$Cf in our experiment was fabricated in April 2013 and total activity was 9000 Bq. the neutron spectrum of $^{252}$Cf conforms to Maxwell distribution:

$$N(E) \propto \sqrt{E} \exp\left(-\frac{E}{T_M}\right) \qquad (1)$$

Where $T_M$ is Maxwell's temperature and it is $(1.453\pm0.017)$ MeV for $^{252}$Cf neutron source. Each spontaneous fission will produce 3.76 neutrons on the average.

Different helium-3 spherical proportional counters with same size have different response even in the same neutron field. Then the inconformity of different proportional counters used need to be evaluated first. Eight helium-3 spherical proportional counters is irradiated by the $^{252}$Cf neutron source respectively with the same distance between counters and $^{252}$Cf. The response of counters was normalized by theoretic neutron flux and the result is shown in Figure 2.

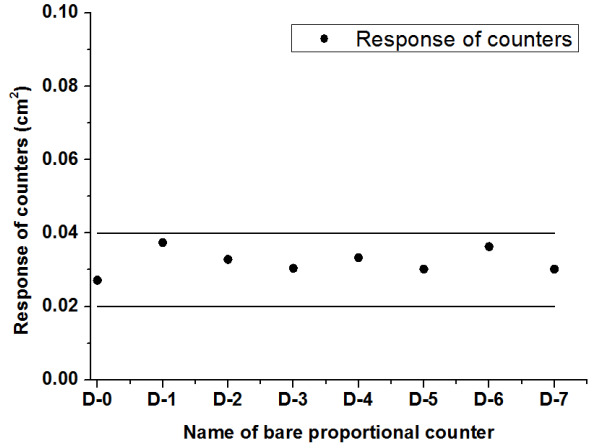

Figure 2 Response of different $^3$He proportional counters

All the response of other counters are corrected based on someone counter, such as D-3 counter. Then the error from counters inconformity can be reduced.

We define the experimental response of each detectors such as DR-50 the following equation.

$$\eta_{exp} = \frac{N_{cps}}{\varphi} = \frac{N_{cps}}{A/(4\pi R^2)} \quad (2)$$

In Equation-2 $\eta_{exp}$ is the experimental response; $N_{cps}$ is the count rate; A is the activity of neutron source; R is the distance between neutron source and detectors; $\varphi$ is the neutron flux at the distance of R.

We define the theoretical response of each detectors the following equation.

$$\eta_{the} = \int \Phi(E) R(E) \quad (3)$$

In Equation-3 $\eta_{the}$ is the theoretical response; $\Phi(E)$ is the theoretical neutron spectrum of $^{252}$Cf (Equation-1); $R(E)$ is response function of the detector.

We define the relative detection efficiency the following equation.

$$\eta = \frac{\eta_{exp}}{\eta_{th}} \quad (4)$$

Then the relative detection efficiency of detectors is the following table.

Table 2 Relative detection efficiency of different detectors

| Name of detectors | Experimental response (cm$^2$) | Theoretical response (cm$^2$) | Relative detection efficiency |
|---|---|---|---|
| DR-0 | 0.064 | 0.009 | 721.50% |
| DR-50 | 0.487 | 1.759 | 27.69% |
| DR-60 | 1.017 | 2.355 | 43.21% |
| DR-70 | 1.147 | 2.707 | 42.37% |
| DR-80 | 1.107 | 2.824 | 39.22% |
| DR-100 | 1.130 | 2.572 | 43.92% |
| DR-120 | 1.014 | 2.023 | 50.14% |
| DR-150 | 0.508 | 1.241 | 40.97% |

Divergence can be found between experimental response and the theoretical, because the following cause:

1) The actual neutron spectrum is not completely Maxwell distribution because neutrons from $^{252}$Cf can be slowed down by material of source.
2) Some fast neutrons will be moderated by materials of source and surroundings which result in more thermal neutrons than the theoretical.

In the table 2, the experimental response of DR-0 is more than the theoretical. The reason is that many fast neutrons are scattered and slowed down by the surroundings. And this phenomenon can be found in the figure 3. Based on genetic algorithm, we unfold the neutron spectrum like the following figure and compare with Maxwell distribution.

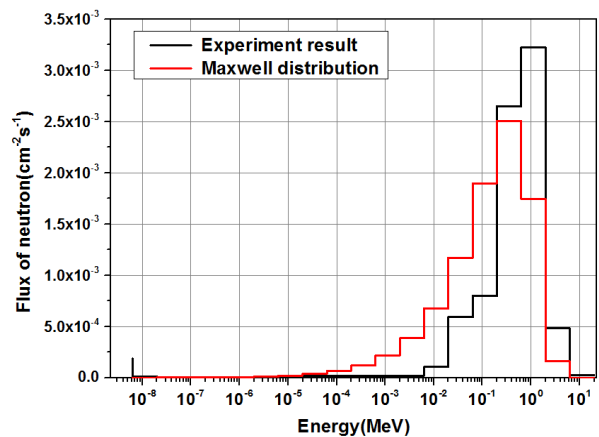

Figure 3 Experimental spectrum of $^{252}$Cf and Maxwell distribution

Total neutron flux from neutron spectrum is $8.20 \times 10^{-3}$ cm$^{-2}$s$^{-1}$. According the distance(R) between neutron source and detectors, we obtain the computational activity of $(156.8 \pm 19.4)$ Bq (uncertainty from uncertainty of relative detection efficiency) and the actual value is $(172.8 \pm 17.3)$ Bq. We can find the difference between experimental spectrum and Maxwell distribution in the figure 3. This is because some neutrons emitted from $^{252}$Cf will be scattered by fabricating materials of neutron source. Then neutron spectrum from source is not completely Maxwell distribution. In addition, neutrons can be scattered and moderated by surroundings, so thermal neutrons can be detected although they are very few in Maxwell distribution.

## 3. Result and discussion

### 3.1 Screening of dataset

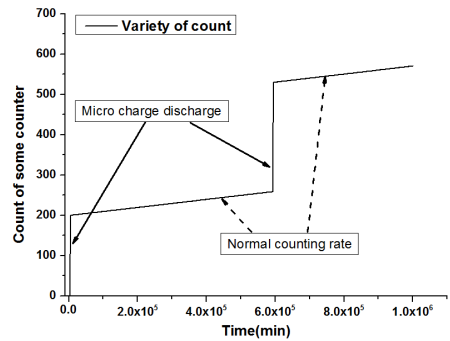

Figure 4 Schematic diagram of screening data

Sometime micro charge discharge appears in the proportional counters and the count will increase sharply shown in the figure 5. Based on the results measured on the ground [11], neutron counting rate at CJPL-1 is less than $10^{-3}$ cps, otherwise the signal may be from micro charge discharge. According to change of counting rate, we screened the valid signal. After screening we obtain the final dataset and get the counting rate by linear fitting.

### 3.2 Material radioactivity of detectors

Because the wall of spherical proportional counter is made up of steel and some radionuclides like $^{238}$U and $^{232}$Th in the steel will produce alphas which can lead to background in the counters. While the content of radionuclide is unknown, we make some assumptions that activity of $^{238}$U and $^{232}$Th are respectively 10 mBq/kg and 2 mBq/kg by the general situation. Based on Monte Carlo simulation, we get the energy deposited spectrum by alphas from the steel. The total counting rate from alphas is $5.13\times10^{-6}$ cps for $^{238}$U decay chain and $9.82\times10^{-7}$ cps for $^{232}$Th decay chain.

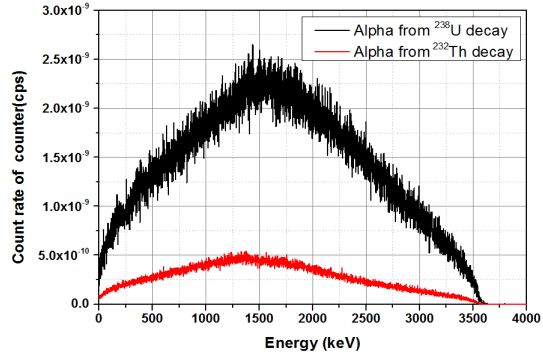

Figure 5 Energy deposited spectrum from steel wall

### 3.3 Discussion of neutron background

From July, 2014 to November, 2015 the detector system was put in CJPL-1 for long term measurement. After screening of dataset and deducting the background from the wall of counters, we get the counting rate of each detector. Then the neutron spectrum is obtained by unfolding code based on genetic algorithm. The result is shown in the following Figure 6.

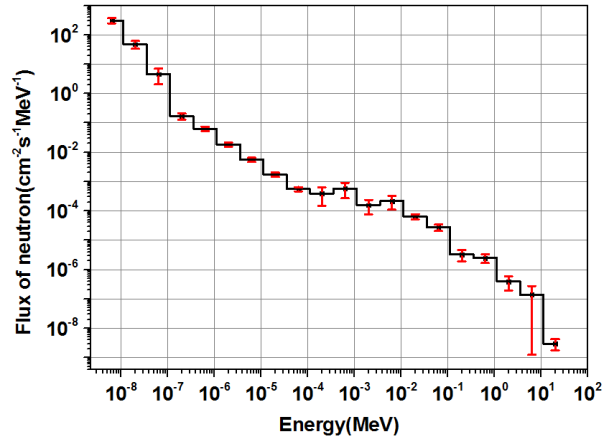

Figure 6 Neutron spectrum measured at CJPL-1

From the unfolded neutron spectrum we get the neutron total flux of $(2.69\pm 1.02)\times 10^{-5}$ cm$^{-2}$s$^{-1}$. The uncertainty is from statistical fluctuation and systematic uncertainty which will be explained in the following. And the flux of thermal neutron (below 0.5 eV) is $(7.03\pm 1.81)\times 10^{-6}$ cm$^{-2}$s$^{-1}$; the flux of fast neutron (above 1 MeV) is $(3.63\pm 2.77)\times 10^{-6}$ cm$^{-2}$s$^{-1}$. The neutron spectrum is similar to the results of Canfranc underground laboratory [13] and agrees with the previous measurement [9]. Table 3 gives some results of neutron flux from different underground laboratories in the world.

Table 3 Neutron flux from different underground labs

| Underground lab | Depth (m.w.e) | Thermal neutron flux ($cm^{-2}s^{-1}$) | Fast neutron flux ($cm^{-2}s^{-1}$) | References |
| --- | --- | --- | --- | --- |
| CPL | 1000 | No data | $(3.00\pm0.02\pm0.05)\times10^{-5}$ | Kim H J et al[14] |
| YangYang | 2000 | $(2.42\pm0.22)\times10^{-5}$ | $8\times10^{-7}$ | Park H et al.[15]; Lee H S et al.[16] |
| Soudan | 2090 | $(0.7\pm0.08\pm0.08)\times10^{-6}$ | No data | Best A et al[17] |
| Canfranc | 2450 | $(1.13\pm0.02)\times10^{-6}$ | $(0.66\pm0.01)\times10^{-6}$ | Jordan D et al.[13] |
| Boulby | 2800 | No data | $(1.72\pm0.61\pm0.38)\times10^{-6}$ | Tziaferi E et al.[18] |
| Gran Sasso | 3600 | $(1.08\pm0.02)\times10^{-6}$ | $(0.23\pm0.07)\times10^{-6}$ | Belli P et al.[19] |
| Modane | 4800 | $(1.6\pm0.1)\times10^{-6}$ | $(4.0\pm1.0)\times10^{-6}$ | Chazal V et al.[20] |
| CJPL | 6720 | $(4.00\pm0.08)\times10^{-6}$ | No data | Zeng Z M et al.[9] |
| CJPL | 6720 | $(7.03\pm1.81)\times10^{-6}$ | $(3.63\pm2.77)\times10^{-6}$ | This study |

From Table 3 we can find the thermal neutron fluxes in the underground labs are almost less than $10^{-5}$ $cm^{-2}s^{-1}$ except Yang Yang. While there is no obvious rule between depth and neutron flux because the neutron flux is dominated by the radioactive components in the surroundings of laboratories when the overburden of rocks is enough deep. And our result of $(7.03\pm1.81)\times10^{-6}$ $cm^{-2}s^{-1}$ is consistent with previous thermal neutron flux of $(4.00\pm0.08)\times10^{-6}$ $cm^{-2}s^{-1}$ to some extent because the location is different between us. The flux of neutron varies much even in the same laboratory but different locations because neutrons can be scattered by surrounding materials to different degrees [21].

Because of unknown background from steel radioactivity, unfolding process and low energy resolution of spectrometer, the obtained neutron spectrum includes some uncertainties. Some known factors are analyzed like the following:

1) Statistical uncertainty

   The statistical fluctuation of counting rate is 5%~15% because of fewer neutrons at the underground laboratory.

2) Systematic uncertainty

   The uncertainty from linear fitting is less 5% and the uncertainty from assumption of steel radioactivity is 9%~27% based on double error estimation for $^{238}$U and $^{232}$Th.

Considering the statistical and systematic uncertainties, we get uncertainties of neutron spectrum in the Figure 6.

## 4. Summary

Neutron background spectrum at CJPL-1 was obtained by a Bonner sphere spectrometer. The total flux of neutron was $(2.69\pm1.02)\times10^{-5}$ cm$^{-2}$s$^{-1}$ and flux of thermal neutron (below 0.5 eV) is $(7.03\pm1.81)\times10^{-6}$ cm$^{-2}$s$^{-1}$. From the neutron spectrum we find the thermal neutrons and intermediate neutrons are in the majority. Our result of thermal neutron flux agrees with the previous and the fast neutron flux is consistent with others such as Modane underground laboratory. The total flux of neutron in CJPL-1 is similar to other underground laboratories and the neutrons are mainly from the rock and concrete of underground laboratory.

## Acknowledgement

This work is partly supported by National Natural Science Foundation of China (No.11175099 &No. 11355001) and Tsinghua University Initiative Scientific Research Program (No.20151080354 & No.2014Z21016).